\documentclass[a4paper,11pt]{article}
\pdfoutput=1
\usepackage{amssymb,amsmath,amsfonts,makeidx,placeins,multirow,tikz}
\usepackage{graphicx,rotate,subcaption,color,slashed,cite,caption,epstopdf,verbatim}
\usepackage{longtable,tabu}
\usepackage{array}
\usepackage[colorlinks=true,
			linkcolor=magenta,
			urlcolor=blue,
			citecolor=blue]{hyperref}

\numberwithin{equation}{section}
\usepackage{wrapfig}
\usepackage{mwe}

\evensidemargin=\oddsidemargin
\title{\color{black}{\bf Bounds on ultralight bosons from the Event Horizon Telescope observation of Sgr A$^*$}}

\author {\bf Akash Kumar Saha,$^{a}$\footnote{akashks@iisc.ac.in} 
	\hspace{4pt} \bf Priyank Parashari,$^{a}$\footnote{ppriyank@iisc.ac.in} 
	\hspace{4pt} \bf Tarak Nath Maity,$^{a}$\footnote{tarak.maity.physics@gmail.com} \\ 
	\hspace{4pt} \bf Abhishek Dubey,$^{a}$\footnote{abhishekd1@iisc.ac.in}
	\hspace{4pt} \bf Subhadip Bouri,$^{b}$\footnote{subhadipb@iisc.ac.in}
\hspace{4pt}  Ranjan Laha$^{a}$\footnote{ranjanlaha@iisc.ac.in}
\\[10pt]
\small\em $^a$Centre for High Energy Physics, Indian Institute of Science, C.\,V.\,Raman Avenue, Bengaluru 560012, India
\\
\small\em $^b$Department of Physics, Indian Institute of Science, C.\,V.\,Raman Avenue, Bengaluru 560012, India
}
\date{}

\begin{document}

\maketitle

\begin{abstract}
Recent observation of Sagittarius A$^*$ (Sgr A$^*$) by the Event Horizon Telescope (EHT) collaboration has uncovered various unanswered questions in black hole (BH) physics. Besides, it may also probe various beyond the Standard Model (BSM) scenarios. One of the most profound possibilities is the search for ultralight bosons (ULBs) using BH superradiance (SR). EHT observations imply that Sgr A$^*$ has a non-zero spin. Using this observation, we derive bounds on the mass of ULBs with purely gravitational interactions. Considering self-interacting ultralight axions, we constrain new regions in the parameter space of decay constant, for a certain spin of Sgr A$^*$. Future observations of various spinning BHs can improve the present constraints on ULBs.    

\end{abstract}
\section{Introduction}
There are ample observational evidences and theoretical problems which indicate the presence of physics beyond the Standard Model (SM) \cite{Craig:2022uua,Asadi:2022njl, Barrow:2022gsu, Adams:2022pbo,  Antypas:2022asj}. In order to address these, various BSM models have been proposed in the literature\,\cite{Lee:2019zbu, Preskill:1982cy, Abbott:1982af}. Ongoing lab-based experiments are leaping forward to probe, and possibly discover, these models\,\cite{Essig:2013lka, Alexander:2016aln, Rappoccio:2018qxp, Essig:2022yzw}. It is possible that the next breakthrough in particle physics may come from astrophysical and/or cosmological observations, notably in scenarios where BSM particles couple very weakly to the SM states. In such circumstances, extreme astrophysical environments might give us clues to answer some of these fundamental questions. For example, our current knowledge of various astrophysical objects like supernovae, neutron stars, BHs, etc.\,\,enable us to place leading constraints on the properties of light relics in large regions of the parameter space\,\cite{Raffelt:1990yz, Raffelt:1996wa, Baryakhtar:2022hbu}.

In this paper, we use spinning BH as a laboratory to test the presence of ULBs. We utilize the phenomenon called SR that happens for a large class of dissipative systems\,\cite{Brito:2015oca, Baryakhtar:2022hbu}. For instance, SR in electromagnetism can be understood by considering an electromagnetic wave incident on an axisymmetric, conducting cylinder which rotates at a constant angular velocity\,\cite{1971JETPL..14..180Z, 1986RaF....29.1008Z}. If the angular velocity of the incident wave is smaller than the rotational velocity of the cylinder ($\Omega$), then after scattering, the wave will extract angular momentum for
\begin{eqnarray}
	\label{SRcondGeneral}
	\frac{\omega_\gamma}{m} < \Omega\,,
\end{eqnarray}
where $\omega_\gamma$ and $m$ are the energy and angular momentum of the incident wave with respect to the rotation axis of the cylinder, respectively. The fact that the outgoing wave has a larger amplitude as compared to the incident one, is known as SR. A similar process can happen for a rotating BH, sometimes known as the Penrose process\,\cite{Penrose:1969pc}. For a BH, the horizon acts as the source of dissipation. In this case, a ULB having a Compton wavelength comparable to the size of the BH horizon may efficiently extract angular momentum from the rotating BH. Note that these ULBs arise from vacuum fluctuations around the BH and can form a bound state with them. These bound states exhibit hydrogen atom-like behavior. These ULBs need not have any cosmic density but must be present in the Lagrangian.

Using SR, BH spin measurements are used to probe ULB particles\,\cite{Arvanitaki:2009fg, Arvanitaki:2010sy, Yoshino:2012kn, Arvanitaki:2014wva, Gruzinov:2016hcq, Baryakhtar:2017ngi, Davoudiasl:2019nlo, Siemonsen:2019ebd, Stott:2020gjj, Baryakhtar:2020gao, Unal:2020jiy, Herdeiro:2021znw, Mehta:2020kwu, Mehta:2021pwf, Ghosh:2021zuf, Du:2022trq, Cannizzaro:2022xyw, Cheng:2022ula, Blas:2020nbs, Blas:2020kaa, Caputo:2021efm, Chung:2021roh, Payne:2021ahy, Roy:2021uye, Baumann:2021fkf, Yuan:2022bem, Chen:2022nbb, Baumann:2022pkl}\footnote{Refs.\,\cite{Day:2019bbh, Chadha-Day:2022inf} have explored SR phenoemena for stars.}. We use the recent imaging of the Milky-Way supermassive black hole (SMBH), Sgr A$^*$, by the EHT collaboration\,\cite{EventHorizonTelescope:2022xnr, EventHorizonTelescope:2022vjs, EventHorizonTelescope:2022wok, EventHorizonTelescope:2022exc, EventHorizonTelescope:2022urf, EventHorizonTelescope:2022xqj} to constrain the properties of ULBs. In particular, EHT has demonstrated that their models with dimensionless spin parameters $0.5$ and $0.94$ have passed all the tests\footnote{Ref.~\cite{Daly:2023axh} has found the dimensionless spin of Sgr A$^*$  0.90 ± 0.06.}\,\cite{EventHorizonTelescope:2022xnr}. We use these two benchmark values of the spin to put constraints on the existence of scalar, vector, and tensor particles. Additionally, we also show the change in our constraints depending upon various possible values of BH spin parameter. Further, self-interaction among the scalar particles may suppress the BH spin-down capability of these ULBs. In this context, we have put the leading constraint on the axion decay constant in some regions of the parameter space using the recent EHT observations. In particular, if Sgr A$^*$ has a spin parameter of 0.94, it constrains a new region of the QCD axion parameter space.

The paper is organized as follows: in sec.\,\ref{sec:SR-review}, we give a brief overview of BH SR. In sec.\,\ref{sec:bound}, we present our constraints. In sec.\,\ref{sec:results}, we briefly discuss our results and associated uncertainties. Then, we conclude in sec.\,\ref{sec:conclusion}.

\section{Brief review of Black Hole Superradiance}
\label{sec:SR-review}
BH SR is a phenomenon of a rotating BH losing its angular momentum and energy due to the existence of a massive bosonic particle\,\cite{Zeldovich:1972spj,Zeldovich:1971a,1971NPhS..229..177P,PhysRevLett.28.994,Starobinsky:1973aij,PhysRevD.22.2323}. Under certain conditions, the ULB will extract angular momentum and energy via superradiant instabilities. As a result, the BH will spin down. Superradiant instabilities lead to an exponential growth of the ULB field around the BH, forming a bound state with the BH, and the resulting configuration is sometimes referred to as the gravitational atom\,\cite{Arvanitaki:2009fg,PhysRevD.22.2323}. The effect of superradiant instabilities is maximal when the Compton wavelength of the particle is comparable to the gravitational radius of the BH. The ratio of the gravitational radius of the BH ($r_g = G_N M_{\rm BH}$) to the ULB's Compton wavelength ($\lambda_c=1/\mu_b$) defines the gravitational fine structure constant, $\alpha \equiv r_g/\lambda_c= G_N M_{\rm BH} \mu_b$, where $\mu_b$, $M_{\rm BH}$, and  $G_N$ are the ULB mass, BH mass, and  Newton's gravitational constant, respectively.  The gravitational fine structure constant determines the efficiency of the superradiant instability\,\cite{Arvanitaki:2014wva,Baumann:2019eav}.  

Growth of the ULB field around the BH occurs only if the angular phase velocity ($\omega_b\sim\mu_b$) of the field is smaller than the angular velocity of the BH event horizon ($\Omega_H$), 
\begin{equation}\label{eq:SRcond}
	\frac{\omega_b}{m} < \Omega_H\,.
\end{equation}
Here $m$ represents the azimuthal angular quantum number, and $\Omega_H$ is defined as 
\begin{equation}
	\Omega_H = \frac{1}{2 r_g}\frac{a_*}{1 + \sqrt{1 - a^{*2}}}\,,
	\label{OmegaH}
\end{equation}
where  $a_*$ is the dimensionless spin parameter defined as $a_* = J_{\rm BH}/ (G_N M_{\rm BH}^2)$ with $J_{\rm BH}$ being the magnitude of the BH angular momentum. The spin parameter lies in the range $0 \leq a_*\leq 1$.  

Besides satisfying Eq.\,\eqref{eq:SRcond},  the energy extraction rate via SR should be faster than the fundamental scale for BH accretion. This can be satisfied if 
\begin{equation}\label{SR:cond2}
	\tau_{\rm SR}<\tau_{\rm BH}\,\,.
\end{equation}
Here $\tau_{\rm BH}$ is the characteristic timescale of the BH, and  the instability timescale for SR, $\tau_{\rm SR}$, is
\begin{eqnarray}\label{eq:tau_Sr}
\tau_{\rm SR}=\frac{\ln N_{\rm max}}{\Gamma^b}\,\,,
\end{eqnarray}
where  $\Gamma^b$  is the superradiant instability growth rate of the ULB cloud, and $N_{\rm max}$ is the maximum occupation number of the cloud after the BH spin downs by $\Delta a_*$. The maximum occupation number is
\begin{eqnarray}\label{eq:Nmax}
	N_{\rm max}=\frac{G_N M_{\rm BH}^2\Delta a_*}{m}\,\,,
\end{eqnarray}
where we conservatively take $\Delta a_*=(1-a_*)$\,\cite{Brito:2015oca,Arvanitaki:2014wva,Arvanitaki:2016qwi}. The superradiant instability growth rate ($\Gamma^b$) is different for scalar\,\cite{Arvanitaki:2010sy,Ternov:1978gq,ZOUROS1979139,Detweiler:1980uk,Dolan:2007mj,Yoshino:2013ofa,Arvanitaki:2014wva,Brito:2015oca,Brito_2015,Arvanitaki:2016qwi,Davoudiasl:2019nlo,Unal:2020jiy,Stott:2020gjj}, vector\,\cite{Rosa:2011my,PhysRevD.86.104017,Pani:2012vp,East:2017mrj,PhysRevD.96.035019,Baumann:2019eav,Davoudiasl:2019nlo,Unal:2020jiy,Stott:2020gjj}, and tensor fields\,\cite{Brito:2013wya,PhysRevLett.124.211101,Unal:2020jiy,Stott:2020gjj}. In the following section, we provide the expressions of $\Gamma^b$ for these three cases and use them to constrain the ULB particle mass using the recent EHT result.

A conservative choice for the characteristic timescale is the Salpeter time. The Salpeter time is the timescale for BH accretion where the compact object is radiating at the Eddington limit, and it is $\tau_{\rm Salpeter} \sim 4.5\times10^7$ yr.  For BHs that are radiating at the super-Eddington limit, the characteristic time is $\sim \tau_{\rm Salpeter}/10$ yr\,\cite{Brito:2015oca}. On the other hand, for accretion onto a BH at a fraction of the Eddington limit, the rate of BH mass increase is
\begin{eqnarray}
	\label{accretion}
	\dot{M}_{\rm acc}=f_{\rm Edd}\dot{M}_{\rm Edd}\sim0.02f_{\rm Edd}\frac{M_{\rm BH}}{10^6M_\odot}\,M_\odot \text{yr$^{-1}$}\, ,
\end{eqnarray}  
where $M_{\rm BH}$ is given in solar mass unit. The above equation assumes the radiative efficiency, $\eta \approx$ 0.1\,\cite{osti_4778507,PhysRevD.89.104059,Brito_2015,Brito:2015oca}. The Eddington ratio for mass fraction, $f_{\rm Edd}$, depends on the detailed properties of the accretion disk surrounding a BH. In case of Sgr A$^*$, $f_{\rm Edd}\sim10^{-9}$\,\cite{Brito:2015oca,Wislocka:2019efh}. For an accreting BH with the accretion rate given by Eq.~\eqref{accretion}, the mass growth is exponential with e-folding time given by a fraction $( 1/{f_{\rm Edd}} )$ of the Salpeter time. Therefore, for Sgr A$^*$, the timescale relevant for gas accretion is $\sim10^{16}$ yr, which is greater than the Hubble time, $\tau_{\rm Hubble} \sim10^{10}$ yr. Thus we take a conservative choice for the characteristic timescale for Sgr A$^*$ in this work and fix it at $\tau_{\rm BH}=5\times10^9$ yr.  A timescale greater than our value will strengthen the constraints on bosonic particles as we will discuss in section (\ref{sec:results}). 

\section{Bounds on bosonic particles}
\label{sec:bound}
In the last section, we gave a summary of BH SR and discussed the necessary conditions for the BH spin depletion via SR. For a given observation of BH parameters, namely, BH mass and spin, we can use the SR conditions (Eqs.\,\eqref{eq:SRcond} and \eqref{SR:cond2}) to put upper and lower bounds on the mass of the ULB particles ($\mu_b$), assuming that depletion in BH spin by $\Delta a_*$ has not occurred due to the SR. In this section, we use the recent observation of Sgr A$^*$ by EHT to constrain the masses of ULB particles with spins 0, 1, and 2. We also consider the case with a self-interacting scalar field and constrain its self-interaction strength using the recent EHT measurement of Sgr A$^*$.

\subsection{Case I: Non-interacting particles}\label{subsec:bound-Ni}
First, we consider the case where ULB particles do not have any interaction other than the gravitational interaction. We refer to this as the non-interacting scenario. For  this case, the superradiant instability rates ($\Gamma^b$) for the ULB particles with spins 0, 1, and 2 have already been discussed in the literature. In the following subsections, we summarize those results briefly.

\subsubsection{Spin-0}\label{subsubsec:bound-Ni_s0}
A massive scalar field ($\Phi$) obeys the Klein-Gordon (KG) equation of motion in a spacetime defined by the metric $g^{\mu \nu}$:
\begin{equation}
	(g^{\mu \nu} \nabla_\mu \nabla_\nu - \mu_S^2)\Phi = 0\,. 
\end{equation}
For BH SR, $g^{\mu \nu}$ will be the Kerr metric of BH under consideration and $\mu_S$ is the mass of the ultralight scalar. Analytical solution to the KG equation is determined using Detweiler's approximation~\cite{Detweiler:1980uk}. This provides the superradiant instability rate for a scalar field as~\cite{Detweiler:1980uk,Baumann:2019eav}				
\begin{equation}
	\Gamma^S_{n \ell m} = 2 \tilde{r}_+ C_{n \ell} \, g_{\ell m}(a_*, \alpha,\omega) \, (m \Omega_H - \omega_{n \ell m})\hskip 1pt \alpha^{4\ell+5}  \, , \label{eqn:ScaRate}
\end{equation}
where 
\begin{align}
	C_{n \ell} &\equiv  \frac{2 ^{4\ell+1} (n+\ell)!}{  n^{2\ell+4} (n-\ell-1)! } \left[ \frac{\ell !}{(2\ell)! (2\ell+1)!} \right]^2  \, , \label{eqn:Cnl} \\
	g_{\ell m}(a_*, \alpha, \omega) &\equiv \prod^{\ell}_{k=1} \left( k^2  \left( 1-a_*^2 \right) + \left( a_* m - 2 r_+ \hskip 1pt  \omega \right)^2  \right)  . \label{eqn:glm}
\end{align}
In the above expressions, $\tilde{r}_{+} = \frac{r_g + \sqrt{r_g^2 + a^2}}{r_g} $ and 
\begin{equation}\label{eq:wnlm}
	\omega_{n \ell m} \approx  \mu_S\left[1- \frac{1}{2} \left(\frac{\alpha}{n+\ell+1}\right)^2\right] \sim \mu_S
\end{equation}	
where $n$, $\ell$, and $m$ are the  principle, orbital angular momentum, and azimuthal angular momentum quantum numbers, respectively.  The dominant growing mode for a scalar field is the dipole mode, $|nlm\rangle =|211\rangle$. We note that the coefficient $C_{n \ell}$ for the dominant mode obtained using Eq.~\eqref{eqn:Cnl} gives $C_{21} = 1/48$, which differs by a factor of 2 with the same of Ref.~\cite{Detweiler:1980uk}. This mismatch was also pointed out in refs.~\cite{Pani:2012bp,Baryakhtar:2017ngi}. Therefore, the growth rate for the dominant mode is
\begin{equation}\label{eq:Gammas_dom}
\Gamma^S_{211} =\frac1{48}a_*\,r_g^8\mu_S^9\,.
\end{equation}
Using the growth rate for the dominant mode and Eqs.~\eqref{SR:cond2} and \eqref{eq:tau_Sr}, we can put an upper limit on the mass of scalar particle demanding that the BH spin is not depleted by SR. The upper limit for the dominant mode is
\begin{equation} \label{eq:dom_up_lim}
	\mu_S<\left(\frac{48\ln N_{\rm max}}{a_*\, r_g^8\tau_{\rm BH}}\right)^{1/9}\, .
\end{equation}
In this work, we use the recent spin measurement of Sgr A$^*$ by the EHT collaboration \cite{EventHorizonTelescope:2022xnr, EventHorizonTelescope:2022vjs, EventHorizonTelescope:2022wok, EventHorizonTelescope:2022exc, EventHorizonTelescope:2022urf, EventHorizonTelescope:2022xqj} and constrain the scalar particle mass.

\subsubsection{Spin-1}\label{subsubsec:bound-Ni_s1}
A massive vector field obeys the Proca equations of motion on a spacetime defined by metric $g^{\mu \nu}$:
\begin{equation}
\nabla_\mu F^{\mu\nu} = \mu_V^2 A^\nu\,,
\end{equation}
where the Proca field  strength $F^{\mu\nu}$ is defined  in terms of the vector potential $A^\mu$ as  $F^{\mu\nu} = \partial^\mu A^\nu - \partial^\nu A^\mu$. The mass of the vector field is denoted by $\mu_V$. Analytical solution to the Proca equation has been obtained in the literature, and it is used to get the SR instability rate for a particle with spin-1. The instability growth rate for a spin-1 particle is given as~\cite{Baumann:2019eav}
\begin{equation}\label{eqn:VecRate}
\Gamma^V_{n\ell jm} = 2 \tilde{r}_+ C_{n \ell j } \, g_{j m}(a_*, \alpha, \omega)  \left( m \Omega_H - \omega_{n \ell j m} \right) \alpha^{2 \ell + 2 j + 5} \, ,  
\end{equation}
where $j$ is the total angular momentum quantum number, the energy levels are denoted by $\omega_{n \ell j m}$ and the coefficients, 
\begin{align}
	C_{n \ell j} \equiv \frac{2^{2\ell + 2 j +1} (n+\ell)!}{n^{2\ell+4}(n-\ell-1)!} &\left[ \frac{(\ell)!}{(\ell + j )!(\ell + j+1)!}\right]^2 \left[ 1 + \frac{ 2  \left( 1+ \ell  - j   \right) \left(1   - \ell  +  j \right)  }{\ell + j}\right]^2 \, , \label{eqn:VectCoeff} \\
	g_{j m}(a_*, \alpha, \omega) &\equiv \prod^{j}_{k=1} \left( k^2  \left( 1-a_*^2 \right) + \left( a_* m - 2 r_+ \hskip 1pt  \omega \right)^2  \right)  . \label{eqn:coeffProd_v}
\end{align}
The growth rate for the vector field is valid for the modes with $j = \ell \pm 1, \ell$. The dominant growing mode for a vector field is the mode with $|nljm\rangle =|1011\rangle$ and the corresponding growth rate is
\begin{equation}
	\Gamma^V_{1011} =4a_*\, r_g^6\mu_V^7\,.
\end{equation}
Using the dominant mode, an upper limit on $\mu_V$ can be obtained as
\begin{equation}\label{eq:muv_dom_lim}
	\mu_V <\left(\frac{\ln N_{\rm max}}{4a_* \, r_g^6\tau_{\rm BH}}\right)^{1/7}\, .
\end{equation}
 
\subsubsection{Spin-2}

\begin{figure}[t]
	\begin{center}
		\includegraphics[height=0.12\textheight, width=1.00\textwidth]{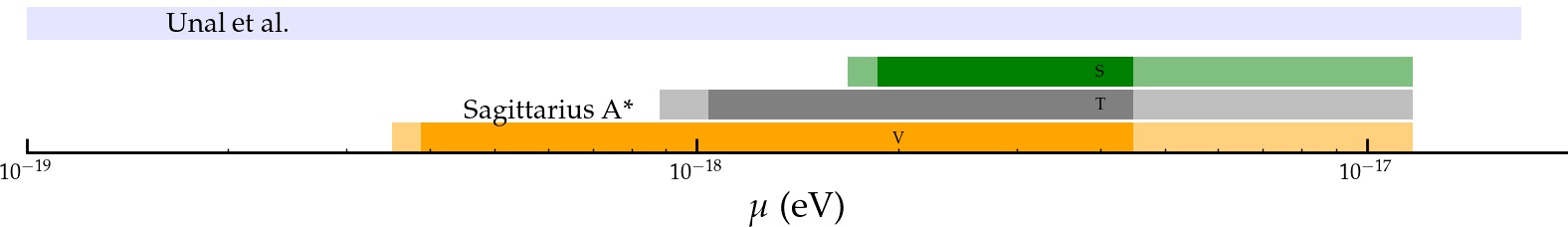}
		\caption{Bounds on non-interacting ULB masses from EHT observation of Sgr A$^*$. Green, orange, and grey shaded regions show the constraints on scalar, vector, and tensor ULBs, respectively. The light and dark shades of these colors indicate the bounds for $a_*=0.94$ and  $a_*=0.5$, respectively. The blue shaded region shows the bound on ULB masses obtained by Ref.\,\cite{Unal:2020jiy}. The systematics in measuring the spin parameters for Sgr A$^*$ and the BHs used in Ref.\,\cite{Unal:2020jiy} are different, and hence our bounds are complementary.}
		\label{fig:bound_svt}
	\end{center}
\end{figure} 
The theories of spin-0 and spin-1 particles are apparent in the SM of particle physics. Particles with higher spin are proposed in several theoretical models \cite{Sorokin:2004ie, Bouatta:2004kk, Sagnotti:2011jdy} and possibly General Relativity is the simplest theory of spin-2 fields. 

The SR instability rate for spin-2 field is\,\cite{Brito:2020lup}
\begin{equation}
\Gamma^T_{n \ell m j} = -C_{j\ell}\frac{{\cal P}_{jm}(a_*)}{{\cal P}_{jm}(0)}\alpha^{2(\ell+j)+5}(\omega_{nlm}-m\Omega_{\rm 
	H})\,, \label{Eq:GammaSpin2}
\end{equation}
where 
\begin{equation}
{\cal 
	P}_{jm}(a_*)=(1+\Delta)\Delta^{2j}\prod_{q=1}^j\left[1+4M_{\rm BH}^2\left(\frac{\omega_{nlm}-m\Omega_{\rm 
		H}}{q\kappa}\right)^2\right]
\end{equation}
is proportional to the BH absorption probability. Here  $\Delta=\sqrt{1-{a_*}^2}$, and $\kappa=\Delta/(1+\Delta)$. The total angular momentum is represented by $j$ and the numerical values of the constant $C_{j\ell}$ for different modes are given in Ref.\,\cite{Brito:2020lup}. Compared to scalar and vector cases, we have more superradiant instability modes with the requirement that the modes have to be nonaxisymmetric. There are two dominant modes, namely dipole ($j=\ell=1$) and quadrupole ($j=2, \ell=0$), however, the numerical values of $C_{j\ell}$ indicate that the quadrupole is the dominant unstable mode. The leading order expression is given by 
\begin{equation}\label{eq:j2l0}
\Gamma_{0022}^T
\simeq \frac{64}{45} {a_*}r_g^8 \mu_T^9\,.
\end{equation}
Using Eq.\,\eqref{eq:j2l0}, and assuming SR has not depleted the BH spin, an upper limit on the mass of the spin-2 particle can be obtained as follows
\begin{equation}
\label{eq:ulspi2}
\mu_T < \left( \frac{45 \, {\rm ln}N_{\rm max}}{64 a_* r_g^8 \tau_{\rm BH}} \right)^{1/9}.
\end{equation}

\begin{figure}[t]
	\begin{center}
		\includegraphics[height=0.35\textheight, width=0.6\textwidth]{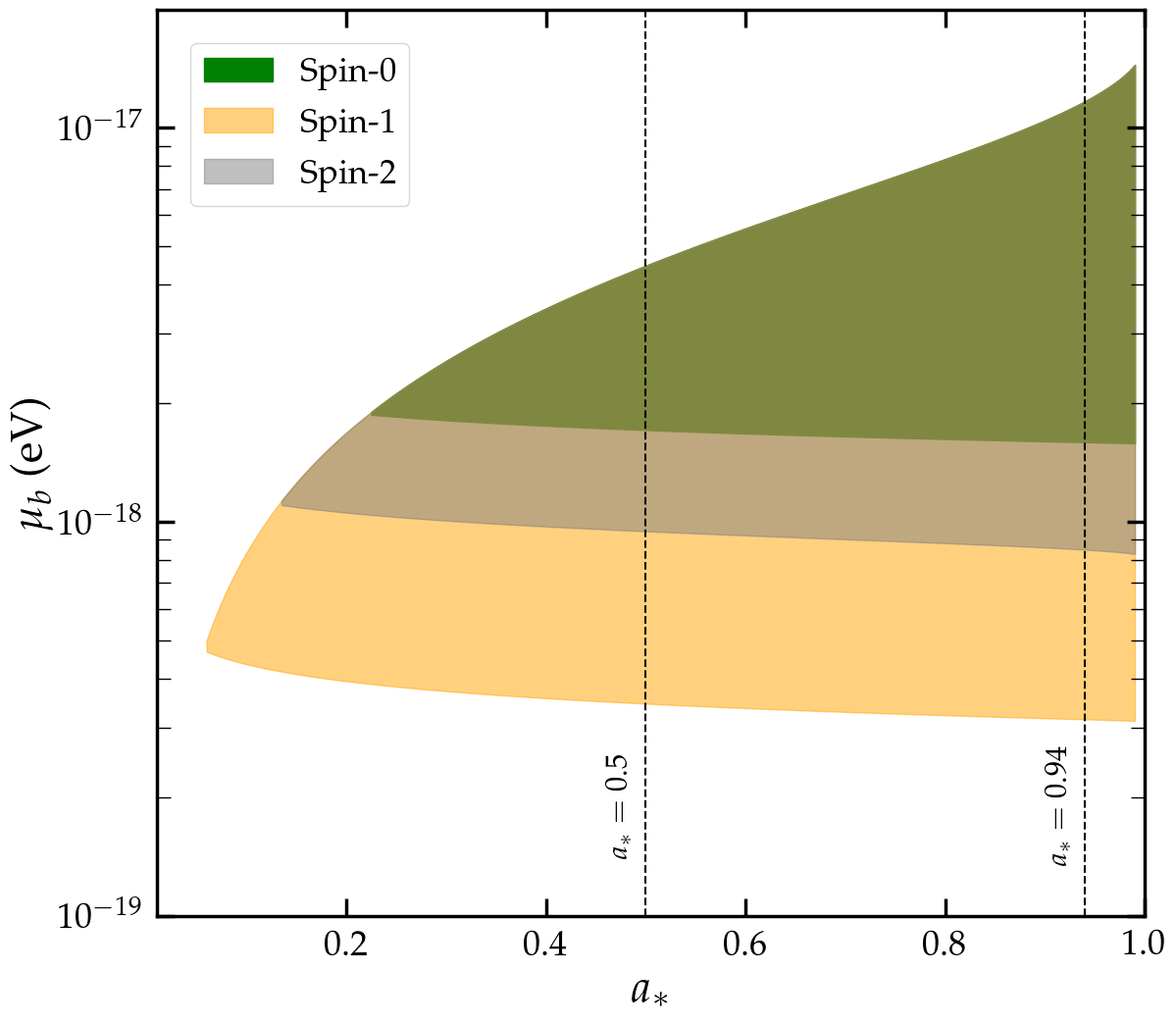}
		\caption{Dependence of the non-interacting ULB mass bounds on the spin of Sgr A$^*$. Green, orange, and grey colors show the bounds for scalar, vector, and tensor ULBs, respectively. The vertical dotted lines show the benchmark spin parameter values tested and confirmed by EHT observation of Sgr A$^*$.}
		\label{fig:bound_svt1}
	\end{center}
\end{figure} 
\subsection{Case II: Self-interacting particles}

The presence of gravitational interaction alone can give rise to BH SR. Many well-motivated theories beyond the SM predict ultralight scalar particles having interaction among themselves and with SM states, for example, the QCD axion \cite{Weinberg:1977ma, Wilczek:1977pj, Peccei:1977hh, Preskill:1982cy}, Kaluza-Klein (KK) modes in string axiverse \cite{Svrcek:2006yi,Arvanitaki:2009fg}, etc. In this section, we explore the effect of scalar self-interaction on BH SR in light of recent EHT results. Particularly our focus will be on axion of mass $\mu_a$ and decay constant $f_a$.

A reasonably strong attractive self-interaction may lead to the collapse of the scalar cloud developed through SR, which sometimes is known as bosenova\,\cite{Arvanitaki:2014wva}. This would essentially suppress the spin-down capability of the SR cloud. 
	The critical number of boson particles required for the collapse of the axion cloud, $N_{\rm BOSE}$, is determined by equating the potential energy and the self-interaction energy of the axions\,\cite{Arvanitaki:2010sy}. This number is given by \cite{Arvanitaki:2010sy, Arvanitaki:2014wva,Brito:2020lup },
\begin{equation}
N_{\rm BOSE}= c \times 10^{94} \frac{n^4}{(r_g \mu_a)^3}\left(\frac{M_{\rm BH}}{10^9 M_{\odot}}\right)^2 \left(\frac{f_a}{M_P}\right)^2,
\end{equation}
with $c \sim 5$, obtained through numerical analysis and $M_P$ being the Planck mass.
Therefore in the presence of self-interaction, SR can spin down the BH only if the SR rate is large
\begin{equation}
\label{eq:SI}
\Gamma^b \tau_{\rm BH} \frac{N_{\rm BOSE}}{N_{\rm max}} > {\rm ln}N_{\rm BOSE}\,. 
\end{equation}
Using the measured BH mass and spin, we obtain an upper bound in the axion mass and decay constant through Eq.\,\eqref{eq:SI}.  For small enough self-coupling, the rate of exchange will not be able to compete with the SR rate, and SR can spin down the BH. The constraint obtained from this consideration would be similar to the bosenova one.
\\
It should be noted that there has been a debate around the formation of bosenova. It has been argued in Ref.\cite{Baryakhtar:2020gao} that due to self-interaction, the energy exchange between different SR levels may lead to a quasi-equilibrium state, which may prevent further growth of SR levels. However, Ref.\cite{Omiya:2022gwu} also suggest possible occurance of bosenova for some cases. In this work, we follow the simpler assumption that bosenova will occur. We leave a detailed anaylsis of bosenova formation for future work.

\begin{table}[h!]
	\centering

	\begin{tabular}{ |p{4cm}|p{3.5cm}|p{3.5cm}| p{3.5cm}| }
		\hline
		Sgr A$^*$ spin &\multicolumn{3}{c|}{Bounds on ULB mass in units of $10^{-19}$ eV}\\ \cline{2-4}
		($a_*$) & Scalar &  Vector  & Tensor \\
		\hline
		$0.94$ & $16.7 \le\mu_{19}\le  117$ & $3.52 \le\mu_{19}\le 117$ & $8.81 \le\mu_{19}\le 117 $ \\
		\hline
		$0.5$ & $18.6 \le\mu_{19}\le  44.7$ & $3.88 \le \mu_{19} \le 44.7$ & $10.4 \le \mu_{19} \le 44.7$ \\
		\hline
	\end{tabular}
	\caption{Constraints on non-interacting spin-0, spin-1 and spin-2 particles from Sgr A$^*$ with $\mu_{19}=\mu_b/(10^{-19}\, {\rm eV})$. We have taken $M_{\rm BH}$= 4$\times10^6M_\odot$ and $\tau_{\rm BH}=5\times10^9$ yr.}
\label{Tab: table}

\end{table}

 \begin{figure}[t]
 	\begin{center}
 		\includegraphics[height=0.5\textheight, width=0.8\textwidth]{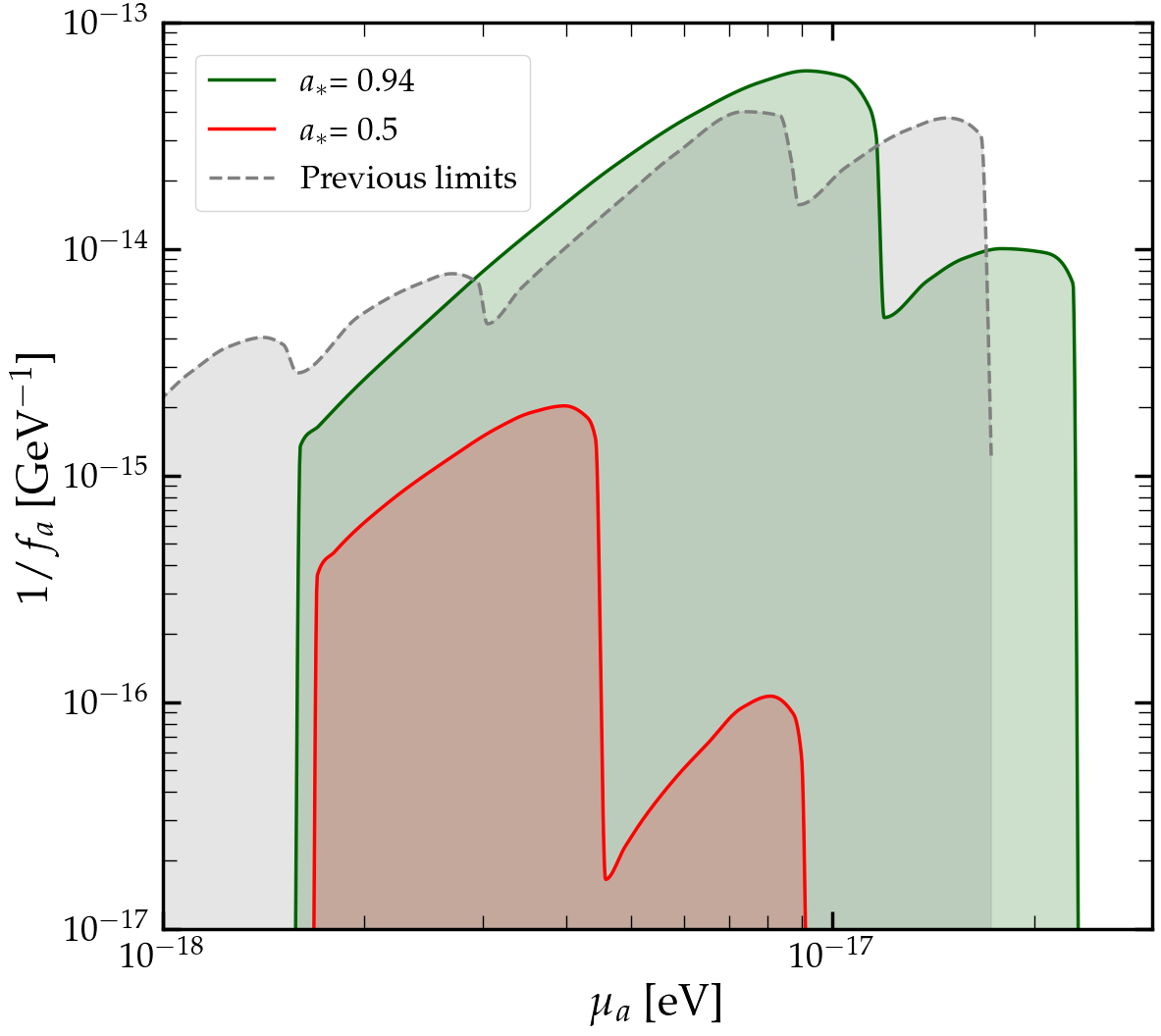}
 		\caption{Bounds on axion decay constant from EHT observation of Sgr A$^*$. The red and green shaded regions show the bounds considering $a_*=0.5$ and $0.94$, respectively. The grey shaded region shows previous limits obtained in\,\cite{Baryakhtar:2020gao,Mehta:2020kwu,Unal:2020jiy,Mehta:2021pwf}}
 		\label{fig:selfInteract}
 	\end{center}
 \end{figure} 
\section{Results and Discussion}
\label{sec:results}


Using the expressions mentioned above, we put conservative bounds on the masses of spin-0, spin-1, and spin-2 ULBs, from EHT observation of Sgr A$^*$. Our results for non-interacting ULBs are shown in Fig.\,\ref{fig:bound_svt}, and given in Table\, \ref{Tab: table}. The obtained bounds are derived for two spin values ($a_*=0.5$ and $0.94$) of Sgr A$^*$. Following Refs.\,\cite{Do:2019txf, abuter2019geometric, reid2020proper}, we take $M_{\rm BH}\, = \, 4\times10^6M_\odot$. From the figure, we see that the constraints on spin-1 and spin-2 ULBs are stronger than in the spin-0 case. This comes from the fact that spin-1 and spin-2 particles interact more with the BH, and as a result, the superradiant instability growth is faster. Thus we get a stronger bound in these two cases.

The constraints from BH SR depend on the BH spin. In Fig.\,\ref{fig:bound_svt1}, we show the dependence of the constraints on the BH spin for Sgr A$^*$ . For example, when $a_*=0.8$, the bounds on scalar, vector, and tensor ULBs are, $\mu_S\in(1.6\times10^{-18},8.3\times10^{-18})$\,eV, $\mu_V\in(3.2\times10^{-19},8.3\times10^{-18})$\,eV, and $\mu_T\in(8.8\times10^{-19},8.3\times10^{-18})$\,eV, respectively. The lower limits of the constrained region mildly depend on the spin parameter, unlike the upper limit. The smallest values of $a_*$ for which we have a constraint on scalar, vector, and tensor ULBs are 0.22, 0.05, and 0.13, respectively. Below these spin values, the lower bounds come out greater than the upper bounds, which implies that there is no common constrained region. We also note that our constraints overlap with those found by Ref.\,\cite{Unal:2020jiy}, where the spin measurements have some uncertainties associated with them. Several proposals have emerged in order to enhance the spin measurement accuracy. Ref.\,\cite{Tamburini:2019vrf} used the twisting of light near a rotating black hole to constrain the spin of M87$^*$ with approximately 5\% accuracy, utilizing current EHT data. Similar explorations can be made for Sgr A$^*$, as indicated in Ref.\,\cite{Chen:2022nbb}. Furthermore, in Ref.\,\cite{Dokuchaev:2023obv} the size of the dark spot observed by the EHT has been used to infer the spin of both Sgr A$^*$ and M87$^*$. Therefore, future observations, which are expected to increase the resolution of black hole images, could potentially improve the accuracy of black hole spin measurements. With these improvements in the SMBH spin measurements, our bounds will become much more robust.

Our limits also depend on the SMBH timescale. However this dependence is rather weak, at most $(\tau_{\rm BH})^{-1/7}$ (for vector particles as shown in Eq.\,(\ref{eq:muv_dom_lim})).  For example, if we change the Sgr A$^*$ timescale from $5\times10^9$ yr (our fiducial value) to $4.5\times10^7$ yr (Salpeter time), the resulting lower bound on bosonic particles will change atmost by a factor of two.

Self-interactions between axions can prohibit the superradiant growth around a BH. Therefore, absence of SR also constrains the axion decay constant\,\cite{Arvanitaki:2014wva, Baryakhtar:2020gao,Unal:2020jiy,Stott:2020gjj,Mehta:2020kwu }. We show this in Fig.\,\ref{fig:selfInteract}. For SR instability rate, we use the dominant mode given in Eq.\,\eqref{eq:Gammas_dom}. In the shaded region of the parameter space, SR can spin down the BH. Light red and green shaded regions correspond to Sgr A$^{\star}$ with spins $0.5$ and $0.94$ respectively. The grey shaded region is the combination of previous bounds in the literature taken from refs.\,\cite{Baryakhtar:2020gao,Mehta:2020kwu,Unal:2020jiy,Mehta:2021pwf}. Interestingly, the latest EHT observation of Sgr A$^*$ allows us to probe some new regions of parameter space for $a_*=0.94$. From Fig.\,\ref{fig:selfInteract} we see that if Sgr A$^*$ has a spin of 0.94 and if it has not been spun down by SR, then this observation probes a new part of the QCD axion parameter space in the mass range $1.73\times10^{-17}$ eV to $2.33\times10^{-17}$ eV. It should be noted that unlike other laboratory experiments, SR spin-down probes smaller coupling. We have presented our results assuming the presence of an axion however, the scenario is also applicable to scalar particles having quartic coupling ($\lambda \leftrightarrow \mu_a^2/f_a^2$) \cite{Arvanitaki:2014wva}. 

Several works have investigated the possible effects of BH environment on the superradiant cloud growth \cite{Arvanitaki:2014wva, Brito:2015oca, Ng:2019jsx, Ng:2020ruv, Cardoso:2020hca, Takahashi:2021eso, Takahashi:2021yhy}. As material from the accretion disk falls into the BH, the spin of the BH increases. This is opposite to the effect of SR, where the ULB cloud spins down the BH. Similarly, for a BH with a small spin, accretion can spin it up and make it subject to superradiant instability growth by satisfying Eq.\,(\ref{eq:SRcond}). If another BH or neutron star merges with the host BH, it can cause substantial perturbation to the superradiant cloud. To estimate all these effects in the case of Sgr A$^*$, we need to know its exact evolution history. We leave this for future work.

\section{Conclusion}
\label{sec:conclusion}

In this work, using the latest observation of Sgr A$^*$ by the EHT collaboration, we constrain ULB masses, assuming that the BH spin has not been depleted via SR. The presence of superradiant instability contradicts the observation of old highly spinning BHs. This in turn puts bounds on purely gravitationally interacting ULBs (see Figs.\,\ref{fig:bound_svt} and \,\ref{fig:bound_svt1}). We also take into consideration the possibility that a non-zero self-interaction can hinder superradiant cloud growth. For ultralight axion this probes a new region of its decay constant (see Fig.\,\ref{fig:selfInteract}). The two most important parameters in this analysis are the BH mass and spin. Though the mass of Sgr A$^*$ is known to considerable accuracy, its spin is yet to be precisely measured. Other works have used various stellar BHs and SMBHs to constrain ULBs using SR\,\cite{Davoudiasl:2019nlo, Unal:2020jiy, Mehta:2020kwu,Ng:2019jsx, Ng:2020ruv, Stott:2020gjj, Arvanitaki:2014wva, Baryakhtar:2020gao}. It is worth noting that near future discoveries of more intermediate-mass BHs can constrain new unexplored regions of parameter space for ULBs \cite{Greene2019IntermediateMassBH, Gais:2022xir, Payne_2022}. Besides, the transitions within superradiant `gravitational atom' cloud give rise to gravitational waves that can be potentially detected by various present and future detectors\cite{Brito:2015oca, Ng:2019jsx, Ng:2020ruv, Arvanitaki:2014wva, Baryakhtar:2020gao}. With the advancement in both theoretical and experimental frontiers, SR can become a very important astrophysical probe to search for new physics in the near future.   

\paragraph*{Acknowledgements\,:}  AKS acknowledges the Ministry of Human Resource Development, Government of India, for financial support through the Prime Minister’s Research Fellowship (PMRF). PP and TNM acknowledge IOE-IISc fellowship program for financial assistance. SB thanks the Council of Scientific and Industrial Research (CSIR), Government of India, for supporting his research under the CSIR Junior/Senior Research Fellowship program through grant no 09/0079(15488)/2022-EMR-I. RL acknowledges discussions with Koushik Chatterjee, Nirupam Roy and Prateek Sharma. RL acknowledges financial support from the Infosys foundation (Bangalore), institute start-up funds, and the Department of Science and Technology (Govt. of India) for the grant SRG/2022/001125, 5, and ISRO-IISc STC for the grant no. ISTC/PHY/RL/499.

The first five authors contributed equally to this work.

\bibliographystyle{JHEP}
\bibliography{ref.bib}  
\end{document}